# Acoustic topological Jackiw-Rebbi states at symmetry broken interfaces

Yifei Xia[1,*], An Chen[1,*], Ting Zhang[1], Jing Yang[1,†], Bin Liang[1,‡], Johan Christensen[2,§], and Jianchun Cheng[1,∥]

[1]Collaborative Innovation Center of Advanced Microstructures and Key Laboratory of Modern Acoustics, MOE, Institute of Acoustics, Department of Physics, Nanjing University, Nanjing 210093, People's Republic of China.

[2]IMDEA Materials Institute, Calle Eric Kandel, 2, 289006, Getafe, Madrid, Spain.

[*]These two authors contributed equally to this work.

Corresponding to: [†]yangj@nju.edu.cn, [‡]liangbin@nju.edu.cn, [§]johan.christensen@imdea.es, and [∥]jccheng@nju.edu.cn.

Topological insulators, a fundamental concept in modern condensed matter physics, support localized states at the interfaces between insulators exhibiting different topological phases, which conventionally rely on explicit symmetry breaking. Here, we propose a mechanism to induce a real-space topological phase transition by spontaneous symmetry breaking, thereby constructing an acoustic metagrating to generate nontrivial Jackiw-Rebbi states characterized by robust imaginary band degeneracy. Our experimental implementation verifies the acoustic delocalized interface state with a constant phase jump, demonstrating enhanced high-directivity topological radiation. We foresee that our findings may spark interest in engineering new acoustic topological devices.

***Introduction.*** – In the vast landscape of condensed matter physics, an intricate arena remains at the forefront of the scenery: topological insulators (TIs) [1-15]. These enigmatic materials defy conventional classification, exhibiting extraordinary properties that stem from the intricate interplay of topology and symmetry [16-24]. TIs, initially explored within electronic systems, have transcended disciplinary boundaries to encompass domains such as acoustics and photonics. In these fields, researchers aim to engineer topologically protected states in metamaterials, leveraging principles of symmetry and topology to confine waves with unprecedented possibilities.

Localized boundary and corner states of topological systems exhibit remarkable robustness against external perturbations, offering avenues for novel applications in energy harvesting, filtering, and advanced waves management in general [25-33]. Amidst the exploration of topological metamaterials, the incorporation of non-Hermitian physics introduces an additional layer of complexity and induces topological transitions, which manifest as intriguing transformations in the topological properties of a system [34-42]. Such topological phase transitions are typically attributed to the explicit symmetry breaking [43,44], exerting a profound influence on the behavior of confined states and presenting fascinating avenues for deeper explorations.

Despite the wealth of research in topological metamaterials, a fascinating frontier remains largely uncharted pertaining to topological robust radiation. In the context of valley-contrasting physics, photonic and acoustic efforts have shown how valley-edge states can be outcoupled from their confining interfaces [45-47]. The topology, typically defined across the entire band within the Brillouin zone, is localized around the valleys, known as the valley Chern number. These valleys, formed by lifting lattice inversion symmetry, hold opposite topological charges, allowing for one-way valley transport, beam splitting, and control of out-coupled valley edge excitations. The Jackiw-Rebbi state is another exotic excitation harnessed for delocalized wave functions enabled by topologically robust control of guided modes [48-51]. Due to the fact that strongly coupled modes cannot degenerate under perturbations at the macroscopic scale, Jackiw-Rebbi states typically depend on explicit rather than spontaneous symmetry breaking [52,53]. This dependence results in a lack of robustness against changes in structural symmetry, thereby imposing strict limitations on its realization in classical wave systems. So far, the observation of the acoustic Jackiw-Rebbi state remains elusive.

In this paper, we uncover an unexplored interplay between topology and symmetry characterized by topological phase transitions arising from spontaneous symmetry breaking in an acoustic symmetric metagrating. Such topological transitions enable the emergence of exceptional points at the interface and the resulting space-dependent Dirac mass, leading to nontrivial Jackiw-Rebbi states to generate topological radiation. This system produces uniquely robust gapless interface states and high Q-factor resonances from imaginary degenerate states. We experimentally confirm the existence of Jackiw-Rebbi states through the observed phase jump of $\pi$, exhibiting bidirectionally enhanced transmission and highly-collimated acoustic beam.

*Acoustic topological metagrating model.* – In the standard model of particle physics, it is well-known that all Dirac fermions acquire their mass through spontaneous symmetry breaking (SSB). When a one-dimensional Dirac field has a spatially dependent mass term arising from SSB, it leads to the generation of Jackiw-Rebbi solitons [54,55]. Here, we design an acoustic topological metagrating that embodies the SSB, which is capable of hosting a Jackiw-Rebbi state at its symmetry-broken interface. The schematic of the acoustic metagrating is illustrated in Figure 1(a). The metagrating consists of rectangular bars periodically arranged alongside a spacer in air. The air region is segmented into four parts, and the velocity potentials in these four regions I, II, III, IV are denoted by $\Phi_1, \Phi_2, \Phi_3$ and $\Phi_4$, respectively. The acoustic impedance of the grating is significantly higher than that of air, thus the boundary of the slit, i.e., the region between the rectangular scatterers, can be considered as a hard boundary. To simplify the guided modes, we neglected higher-order decaying waves in the slit region. Consequently, the acoustic field in the four regions can be expressed as follows:

$$\begin{cases} \Phi_1 = \sum_{n=-\infty}^{\infty} A_n \exp\left(j(k_x + qn)x + jk_y(y-d)\right), y > d \\ \Phi_2 = U(x)[B_1 \exp(jk_n y) + B_2 \exp(-jk_n y)], 0 < y < d \\ \Phi_3 = U(x)[C_1 \exp(jk_n y) + C_2 \exp(-jk_n y)], -d < y < 0 \\ \Phi_4 = \sum_{n=-\infty}^{\infty} D_n \exp\left(j(k_x + qn)x + jk_y(y+d)\right), y < -d, \end{cases} \quad (1)$$

where $k_x$ and $k_y$ represent the wavevector components in the $x$ and $y$ directions, respectively, and satisfy the condition $(k_x + 2\pi n/\Lambda)^2 + k_y^2 = k_0^2$. The term $k_n$ denotes the imaginary wavevector component of the guided mode inside the slit along the $y$-direction. The Bragg wavevector is represented as $q = 2\pi/\Lambda$. Based on the rigorous coupled mode theory, we utilize the continuity conditions at the interface between these four regions to derive the coupling among guided modes. To support the SSB, it is essential to facilitate weak coupling between the guided modes on either side of the slit, which is enabled by the spacer composed of low-impedance material. Given that the thickness of the spacer is significantly smaller than that of the grating, it is appropriate to characterize it as a continuity condition described by an effective transfer impedance calculated as $Z_{\text{eff}} = (Z_u - Z_l)/Z_{\text{air}}$, accounting for the discrepancy in acoustic impedance between the upper and lower sides. The characteristics of the spacer enable us to approximate that the velocities on both sides are equal, while the velocity potentials are different. Therefore, the continuity condition near the spacer can be represented as [56]:

$$\begin{cases} \dfrac{jZ_{\text{eff}}}{k_n}\dfrac{\partial \Phi_2}{\partial y} = \Phi_2 - \Phi_3 \\ \dfrac{\partial \Phi_2}{\partial y} = \dfrac{\partial \Phi_3}{\partial y}. \end{cases} \quad (2)$$

Combining all continuity conditions among the four regions, we obtain the resonance condition for guided modes $\psi = (B_1, C_2)^T$:

$$\begin{pmatrix} Z_{\text{eff}}(R_1 - 1)/2 - R_1 & 1 \\ 1 & Z_{\text{eff}}(R_2 - 1)/2 - R_2 \end{pmatrix} \psi = 0. \quad (3)$$

where $R_1 = B_2/B_1, R_2 = C_1/C_2$ depends on the impedance properties of the metagrating surface and represents the proportionality relationship between the guided modes propagating in opposite directions. When the metagrating is periodic and infinite, $R_1 = R_2 = R$ holds for all slits of the metagrating, indicating the reflection coefficient of the metagrating and determining the Dirac mass. It is possible to adequately describe this system using a Hamiltonian to characterize the band structures for two distinct symmetries. We consider approximating $R$ near a certain purely real wavenumber $k_0$ by a first-order Taylor expansion. The equation (3) can then be simplified as:

$$H\psi = \Delta k_0 \frac{\partial R}{\partial k_0} \psi, \quad (4)$$

where the Hamiltonian can be expressed as:

$$H = \begin{pmatrix} R - \dfrac{Z_{\text{eff}}}{Z_{\text{eff}} - 2} & \dfrac{2}{Z_{\text{eff}} - 2} \\ \dfrac{2}{Z_{\text{eff}} - 2} & R - \dfrac{Z_{\text{eff}}}{Z_{\text{eff}} - 2} \end{pmatrix}. \quad (5)$$

The even and odd eigenvectors and the corresponding eigenvalues are given by:

$$\psi^{\pm} = \begin{pmatrix} \pm 1 \\ 1 \end{pmatrix}, \lambda^{\pm} = R - \frac{\pm 2 - Z_{\text{eff}}}{2 - Z_{\text{eff}}}. \quad (6)$$

We emphasize that this Hamiltonian commutes with the parity operator, indicating that the metagrating exhibits invariance under spatial inversion transformation. Therefore, we compare the slit region to a symmetric double-well potential. The impedance boundary conditions on the upper and lower sides can be acoustically regarded as semi-infinite potential barriers, while the spacer in the middle serves as a narrow potential barrier. Such an equivalence is verified by the similarity between the eigenmodes of our model and that of a double-well potential model as shown in Figure 1(b). Since the potential function is symmetric, the eigenmodes under periodic boundary conditions must be either symmetric or antisymmetric, as shown in Figure 1(c). There exists a bandgap between the eigenmodes with even parity and those with odd parity, which increases with the decrease in effective transfer impedance, as depicted in Figure 1(d). This indicates that the size of the bandgap can be directly controlled by adjusting the height of the acoustic potential barrier, i.e., by altering the transfer impedance of the spacer.

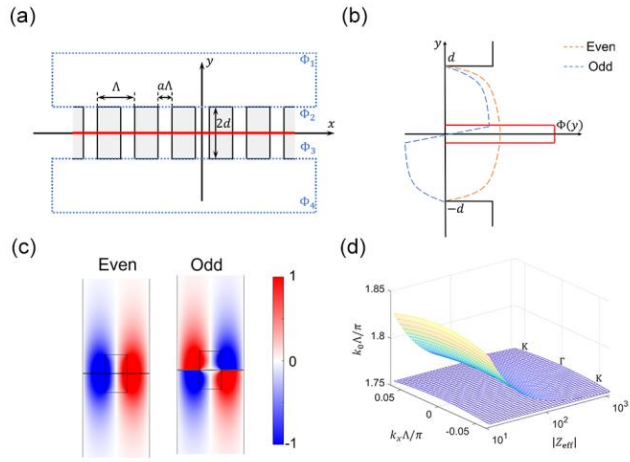

FIG.1. (a) Schematic of the acoustic topological metagrating. The spacer of negligible thickness (colored in red) and the one-dimensional binary lattice (colored in grey) are aligned with the $x$-axis. The period of metagrating is designated as $\Lambda$, where the rectangular scattering unit possess a width of $(1-a)\Lambda$, with $0 < a < 1$, and a height of $2d$. (b) Schematic of an effective double-well potential model of the proposed system shown in (a). The upper and lower sides of the system are characterized by impedance boundary conditions, acting as semi-infinite potential barriers. The spacer positioned between them functions as a narrow potential barrier, highlighted in red. The dashed lines represent the wave functions of the two eigenmodes with different types of parity. (c) Normalized acoustic pressure distributions of the even and odd eigenmodes obtained through simulations. (d) Spectral features of even and odd eigenmodes on $k_x - k_0$ plane. The bandgap between the even and odd eigenmodes decreases with the increase in the transfer impedance of the spacer.

***Real-space topological phase transition induced by SSB.*** – For this double-well potential model, explicit parity symmetry remains unbroken. However, the symmetry of the wave function may not necessarily align with that of the potential. Due to the high acoustic potential barriers, the coupling between guided modes on either side of the spacer is sufficiently weak, making them susceptible to perturbations. Specifically, the symmetry of guided modes in one slit can be influenced by the guided modes of adjacent slits, which leads to SSB at the interfaces of different metagratings when eigenmodes degenerate. In this case, the reflection coefficient $R_1$ and $R_2$ on the upper and lower sides of the metagrating are different. As a result, the interface Hamiltonian breaks parity symmetry, which can be represented as:

$$H' = m\sigma_z + \frac{2}{Z_{\text{eff}} - 2}\sigma_x, \tag{7}$$

$$m = (R_1 - R_2)/2. \tag{8}$$

Equation (7) conforms to the form of the Dirac Hamiltonian. In a single periodic metagrating, the Dirac mass $m$ remains zero due to the unbroken symmetry. However, at the interface between two different metagratings, the Dirac mass is not equal to zero and varies with position. When the Dirac mass reaches specific values, the interface exhibits degenerate eigenmodes at two different exceptional points as depicted in Figure 2(a) and (b), where the Dirac masses $m_{1,2}$ and eigenvectors $\psi_{1,2}^{\pm}$ are expressed as follows:

$$m_1 = \frac{2j}{Z_{\text{eff}} - 2}, \psi_1^+ = \psi_1^- = \begin{pmatrix} j \\ 1 \end{pmatrix}, \tag{9}$$

$$m_2 = \frac{-2j}{Z_{\text{eff}} - 2}, \psi_2^+ = \psi_2^- = \begin{pmatrix} j \\ -1 \end{pmatrix}, \tag{10}$$

where the Dirac masses $m_1$ and $m_2$ are approximately a pair of opposite real numbers. Because Dirac masses $m_{1,2}$ are sufficiently small with the transfer impedance $|Z_{\text{eff}}| \gg 1$, these two exceptional points can be spontaneously induced under intrinsic perturbations. To generate the corresponding degenerate eigenmodes, we arrange two different metagratings with different types of parity along the $x$-axis, as depicted in Figure 2(c). The two exceptional points are positioned on both sides of the interface and close to it, leading to a change in the sign of the Dirac mass at the interface, thereby generating Jackiw-Rebbi states, as illustrated in Figure 2(d). This variation in Dirac mass is independent of the wavevectors $k_x$, indicating that it is caused by real-space topology rather than band topology, which manifests as the absence of the band inversion [57]. Here, we define the real-space topological invariant $\text{sgn}(R_1 - R_2)$ to characterize the topological phase transition occurring in this system, which manifests as a phase shift in the reflected wave. Since the phase distribution of the delocalized waves on the surface is consistent with the form of the eigenvectors, we can further deduce that the phases of the reflected waves on both sides of the metagrating interface are $\phi_{r1} = 0$ and $\phi_{r2} = \pi$, while the phase of the transmitted acoustic beam is $\phi_t = \pi/2$. To verify this deduction, we numerically compute the acoustic near-field surrounding the metagrating as shown in Figure 2(e). Here, the parameters $a$ for the left and right metagratings are chosen as $[a_L, a_R] = [0.71, 0.67]$, while the effective transfer impedance $Z_{\text{eff}}$ are set to $[|Z_L|, |Z_R|] = [1200, 120]$. This configuration ensures the occurrence of a real-space topological transition [57]. The simulations substantiate the presence of localized standing waves near the interface along the surface of the metagrating. Moreover, due to the real-space topological transition, there exists a phase jump of $\pi$ for the reflected wave near the interface, as illustrated in Figure 2(f). The observation of phase discontinuities validates the proposed real-space topology. Notably, due to the mirror symmetry of the metagrating, it exhibits bidirectionally enhanced transmission characteristics, with an approximately 20 dB acoustic enhancement observed at the interface, where the intensity exponentially decays towards both sides. The decaying rate is consistent with the results obtained from theoretical calculations. Concurrently, a leaky acoustic beam with a half-width of approximately $2\Lambda$ and a divergence angle of 0.1 rad is observed, emanating in a direction approximately perpendicular to the metagrating surface.

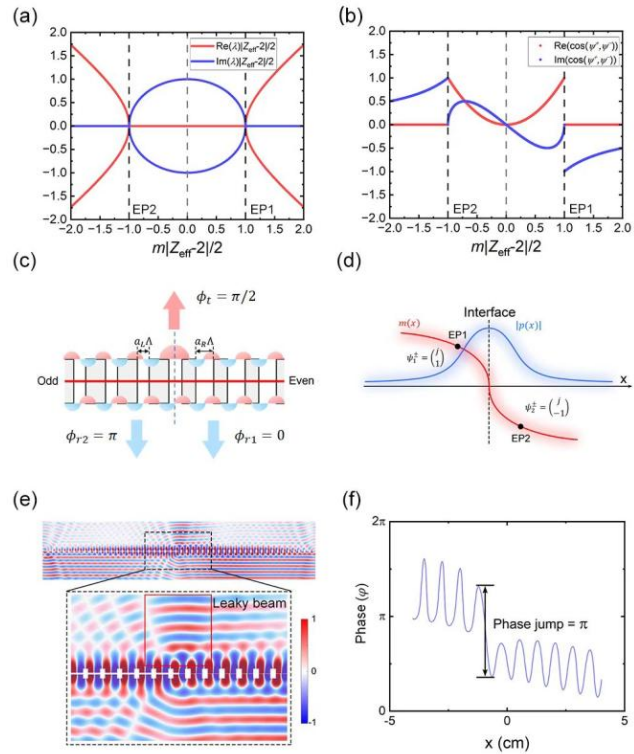

FIG.2. (a) Variation of the eigenvalues of the interface Hamiltonian with respect to the Dirac mass. (b) Variation of the angle between the two eigenvectors of the interface Hamiltonian with respect to the Dirac mass. (c) Schematic of the concatenation of two acoustic metagratings with different types of parity. (d) Degenerate eigenmodes yield a sign-change in the Dirac mass across the interface, forming a Dirac mass domain-wall which results in an acoustic Jackiw-Rebbi state. (e) Computed acoustic pressure distribution of the Jackiw-Rebbi states along the metagrating. A leaky acoustic beam with a half-width of approximately $2\Lambda$ is observed (f) Phase distribution of the reflected wave along the $x$-axis. Due to the real-space topological transition, there exists a phase jump of $\pi$ for the reflected wave near the interface.

To guarantee the emergence of Jackiw-Rebbi states, it is crucial to ensure that eigenmodes with different types of parity occur at the same frequency. As illustrated in Figure 3(a), when the bands of the two metagratings intersect, resonant peaks with high Q-factor appear in the transmission spectrum near the degenerate point, indicating the presence of Jackiw-Rebbi states at the intersection frequency. Conversely, when the bands of the two metagratings do not cross, resonant acoustic fields are localized in one of the gratings and maintain a lower transmission coefficient, thus preventing the generation of Jackiw-Rebbi states. To induce band crossing, the curvatures of the two bands must differ, primarily determined by the slit width $a$. As illustrated in Figure 3(b), when $a \geq 2/3$, the band curvature is relatively small, indicating a flat band. Instead, when $a < 2/3$, the band curvature rapidly increases with decreasing $a$, defining a steep band. By adjusting the structural parameters of the metagratings, we achieve flat band and steep band, ensuring that the bands do not intersect in the real wavevector domain, as depicted in Figure 3(c). Meanwhile, two bands with different types of parity intersect at a specific point in the purely imaginary wavevector domain, as illustrated in Figure 3(d), giving rise to a Jackiw-Rebbi state at the interface. This imaginary band degeneracy persists even after breaking the mirror symmetry of the structure, thereby ensuring the emergence of SSB, which exhibits a unique robustness compared to those relying on explicit symmetry breaking [57].

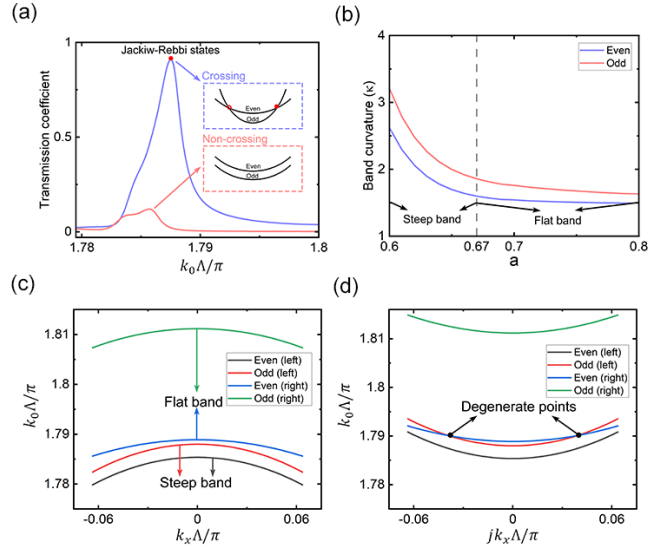

FIG.3. (a) Transmission spectrum of the metagrating. In the presence of band crossing, resonant peaks appear near the intersection frequency, corresponding to the Jackiw-Rebbi states. In the absence of band crossing, no resonant peaks are observed, indicating the absence of Jackiw-Rebbi states. (b) Band curvature varies with the parameter $a$, where $a \geq 2/3$ and $a < 2/3$ correspond to flat bands and steep bands, respectively. (c) Spectral features of two metagratings on the left and right side in the real wavevector domain. Bands do not intersect in the real wavevector domain. (d) Spectral features of two metagratings in the purely imaginary wavevector domain. Two bands with different types of parity intersect at a specific point, which gives rise to a Jackiw-Rebbi state at the interface.

*Experimental realization.* – To experimentally observe Jackiw-Rebbi states, we introduce planar-type spacers and consequently generate acoustic Jackiw-Rebbi states in a two-dimensional metagrating as shown in Figure 4(a). This metagrating is composed of two lattice plates containing periodic square apertures and a spacer sandwiched between the two lattice plates [57]. By adjusting the size of the apertures on the lattice plates, we design a symmetry-broken interface, which divides the lattice plate into two parts. Figure 4(b) presents the metagrating and measuring system employed in our experiment. The experimental results show that the acoustic Jackiw-Rebbi states have a center frequency of 42,200 Hz, a relative bandwidth of 1.6%, and a peak transmittance of 0.86, as illustrated in Figure 4(c). Measurements of the acoustic intensity at various positions were conducted at different heights above the metagrating surface, and the results are presented in Figure 4(d)-(f). The measured acoustic intensity distributions, phases of reflected wave and leaky acoustic beam agree well with the result obtained from simulations. The experimental results demonstrate that the acoustic intensity near the interface reaches its maximum and exponentially decays towards both sides. Moreover, the observed acoustic enhancement effect and decay rate are in accordance with theoretical calculations. These observations of acoustic intensity and phase distribution indicate the occurrence of Jackiw-Rebbi states and the topological phase transition.

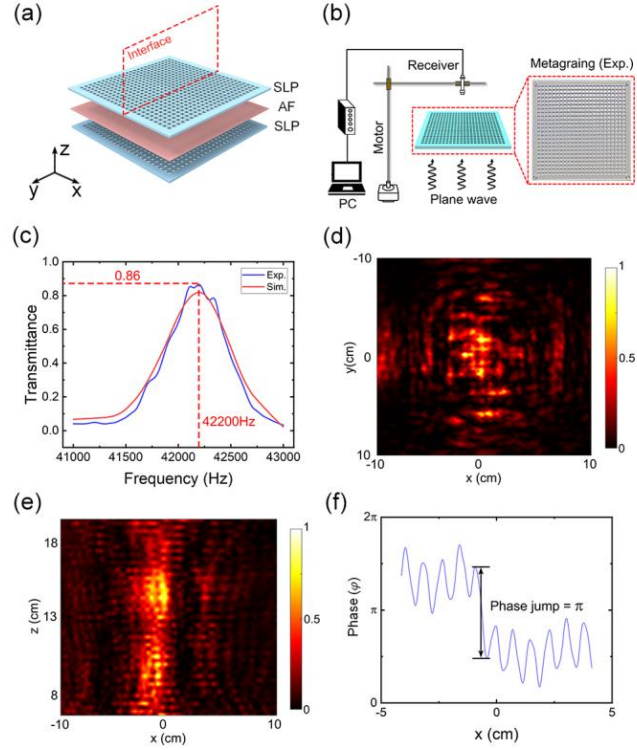

FIG. 4. (a) Schematic of the two-dimensional acoustic metagrating. The metagrating consists of two steel lattice plates containing periodic square apertures and a sheet of aluminum foil sandwiched between the two lattice plates. In the experiment, the metagrating is placed on the x-y plane, the y-z plane is the symmetry broken interface of the acoustic grating. (b) Experimental setup for validating acoustic Jackiw-Rebbi states. An ultrasonic transmitter array serves as the plane wave source, positioned centrally beneath the metagrating. Through the control of ultrasonic receivers using motors, we can measure the acoustic intensity of transmitted or reflected waves at distinct spatial positions. The inset shows the photo of the metagrating. (c) Transmission spectrum obtained from both theoretical calculations and experimental measurements. (d) Acoustic intensity distributions measured on the $x$-$y$ plane at the height of 5cm above the metagrating surface. (e) Leaky acoustic beam on the $x$-$z$ plane obtained from far field measurements. (f) Phase distribution of the reflected wave along the $x$-axis measured beneath the metagrating.

*Conclusions*. – In summary, we propose an acoustic topological metagrating model that enables spontaneous parity symmetry breaking attributed to the degeneracy induced by perturbations, supported by the sufficiently high acoustic potential barrier. Based on this distinctive topological mechanism, we realize a real-space phase transition at the symmetry-broken interface. We theoretically predict and experimentally validate the existence of the Jackiw-Rebbi state along the metagrating, demonstrating its robustness through imaginary band degeneracy. Particularly, we observe the topological radiation manifested as a transmitted acoustic beam with a 20 dB enhancement and a 0.1 rad divergence angle. Our studies underline the importance of nontrivial structure design for exotic topological acoustics that go beyond traditional boundary confinements and hold potential for extension to higher-order topology.

**Acknowledgments:**

This work was supported by the National Key R&D Program of China (Grant Nos. 2022YFA1404402 to B. L.), the National Natural Science Foundation of China (Grant Nos. 12174190 to J. Y.), High-Performance Computing Center of Collaborative Innovation Center of Advanced Microstructures and a Project Funded by the Priority Academic Program Development of Jiangsu Higher Education Institutions. J.C. acknowledges support from the Spanish Ministry of Science and Innovation through a Consolidación Investigadora grant (CNS2022-135706).

**References**

[1] X. Zhang, F. Zangeneh-Nejad, Z.-G. Chen, M.-H. Lu, and J. Christensen, Nature 618, 687 (2023).

[2] Y. Qi, C. Qiu, M. Xiao, H. He, M. Ke, and Z. Liu, Phys. Rev. Lett. 124, 206601 (2020).

[3] Q. Wei, X. Zhang, W. Deng, J. Lu, X. Huang, M. Yan, G. Chen, Z. Liu, and S. Jia, Phys. Rev. Lett. 127, 255501 (2021).

[4] M. Wang, Q. Ma, S. Liu, R.-Y. Zhang, L. Zhang, M. Ke, Z. Liu, and C. T. Chan, Nat. Commun. 13, 5916 (2022).

[5] Y. Yang, J. Lu, M. Yan, X. Huang, W. Deng, and Z. Liu, Phys. Rev. Lett. 126, 156801 (2021).

[6] X. Wu, H. Fan, T. Liu, Z. Gu, R.-Y. Zhang, J. Zhu, and X. Zhang, Nat. Commun. 13, 6120 (2022).

[7] Y. Ding, Y. Peng, Y. Zhu, X. Fan, J. Yang, B. Liang, X. Zhu, X. Wan, and J. Cheng, Phys. Rev. Lett. 122, 014302 (2019).

[8] D. Wang, Y. Deng, J. Ji, M. Oudich, W. A. Benalcazar, G. Ma, and Y. Jing, Phys. Rev. Lett. 131, 157201 (2023).

[9] J.-J. Liu, Z.-W. Li, Z.-G. Chen, W. Tang, A. Chen, B. Liang, G. Ma, and J.-C. Cheng, Phys. Rev. Lett. 129, 084301 (2022).

[10] C. He, H.-S. Lai, B. He, S.-Y. Yu, X. Xu, M.-H. Lu, and Y.-F. Chen, Nat. Commun. 11, 2318 (2020).

[11] H. Xue, D. Jia, Y. Ge, Y.-J. Guan, Q. Wang, S.-Q. Yuan, H.-X. Sun, Y. D. Chong, and B. Zhang, Phys. Rev. Lett. 127, 214301 (2021).

[12] J. Du, T. Li, X. Fan, Q. Zhang, and C. Qiu, Phys. Rev. Lett. 128, 224301 (2022)

[13] F. Gao, Y.-G. Peng, X. Xiang, X. Ni, C. Zheng, S. Yves, X.-F. Zhu, and A. Alù, Adv. Mater. 2312421 (2024).

[14] S. A. Hassani Gangaraj, C. Valagiannopoulos, and F. Monticone, Phys. Rev. Res. 2, 023180 (2020).

[15] P. Gehring, C. Merckling, R. Meng, V. Fonck, B. Raes, M. Houssa, J. Van de Vondel, and S. De Gendt, APL Mater. 11, 111116 (2023).

[16] L.-Y. Zheng and J. Christensen, Phys. Rev. Lett. 127, 156401 (2021).

[17] F. Gao, X. Xiang, Y.-G. Peng, X. Ni, Q.-L. Sun, S. Yves, X.-F. Zhu, and A. Alù, Nat. Commun. 14, 8162 (2023).

[18] G. Ma, M. Xiao, and C. T. Chan, Nat. Rev. Phys. 1, 281 (2019).

[19] Z. Wang, X. Wang, Z. Hu, D. Bongiovanni, D. Jukić, L. Tang, D. Song, R. Morandotti, Z. Chen, and H. Buljan, Nat. Phys. 19, 992 (2023).

[20] I. Zeljkovic, Y. Okada, M. Serbyn, R. Sankar, D. Walkup, W. Zhou, J. Liu, G. Chang, Y. J. Wang, M. Z. Hasan, F. Chou, H. Lin, A. Bansil, L. Fu, and V. Madhavan, Nat. Mater. 14, 318 (2015).

[21] L. Lu, C. Fang, L. Fu, S. G. Johnson, J. D. Joannopoulos, and M. Soljačić, Nat. Phys. 12, 337 (2016).

[22] Y. Wang, B. Liang, and J. Cheng, Sci. China. Ser. G: Phys. Mech. Astron. 67, 224311 (2024).

[23] X. Ni, M. Weiner, A. Alù, and A. B. Khanikaev, Nat. Mater. 18, 113 (2019).

[24] L. J. Maczewsky, B. Höckendorf, M. Kremer, T. Biesenthal, M. Heinrich, A. Alvermann, H. Fehske, and A. Szameit, Nat. Mater. 19, 855 (2020).

[25] C.-W. Chen, R. Chaunsali, J. Christensen, G. Theocharis, and J. Yang, Commun. Mater. 2, 1 (2021).

[26] W. Wang, X. Wang, and G. Ma, Phys. Rev. Lett. 129, 264301 (2022).

[27] X. Wu, Y. Meng, Y. Hao, R.-Y. Zhang, J. Li, and X. Zhang, Phys. Rev. Lett. 126, 226802 (2021).

[28] H. Fan, B. Xia, L. Tong, S. Zheng, and D. Yu, Phys. Rev. Lett. 122, 204301 (2019).

[29] A. El Hassan, F. K. Kunst, A. Moritz, G. Andler, E. J. Bergholtz, and M. Bourennane, Nat. Photonics 13, 697 (2019).

[30] H.-R. Kim, M.-S. Hwang, D. Smirnova, K.-Y. Jeong, Y. Kivshar, and H.-G. Park, Nat. Commun. 11, 5758 (2020).

[31] H.-S. Lai, H. Chen, C.-H. Xia, S.-Y. Yu, C. He, and Y.-F. Chen, Research 6, 0235 (2023).

[32] E. W. Wang, D. Sell, T. Phan, and J. A. Fan, Opt. Mater. Express, OME 9, 469 (2019).

[33] A. Sarsen and C. Valagiannopoulos, Phys. Rev. B 99, 115304 (2019).

[34] A. Fritzsche, T. Biesenthal, L. J. Maczewsky, K. Becker, M. Ehrhardt, M. Heinrich, R. Thomale, Y. N. Joglekar, and A. Szameit, Nat. Mater. 23, 377 (2024).

[35] Y. G. N. Liu, P. S. Jung, M. Parto, D. N. Christodoulides, and M. Khajavikhan, Nat. Phys. 17, 704 (2021).

[36] L. Li, C. H. Lee, and J. Gong, Phys. Rev. Lett. 124, 250402 (2020).

[37] Y. Li, C. Liang, C. Wang, C. Lu, and Y.-C. Liu, Phys. Rev. Lett. 128, 223903 (2022).

[38] P. Gao, M. Willatzen, and J. Christensen, Phys. Rev. Lett. 125, 206402 (2020).

[39] B. Hu, Z. Zhang, H. Zhang, L. Zheng, W. Xiong, Z. Yue, X. Wang, J. Xu, Y. Cheng, X. Liu, and J. Christensen, Nature 597, 655 (2021).

[40] L. Zhang, Y. Yang, Y. Ge, Y.-J. Guan, Q. Chen, Q. Yan, F. Chen, R. Xi, Y. Li, D. Jia, S.-Q. Yuan, H.-X. Sun, H. Chen, and B. Zhang, Nat. Commun. 12, 6297 (2021).

[41] X.-W. Luo and C. Zhang, Phys. Rev. Lett. 123, 073601 (2019).

[42] S. A. Hassani Gangaraj and F. Monticone, Phys. Rev. Lett. 121, 093901 (2018).

[43] T. Suchanek, K. Kroy, and S. A. M. Loos, Phys. Rev. E 108, 064123 (2023).

[44] T. Dai, Y. Ao, J. Mao, Y. Yang, Y. Zheng, C. Zhai, Y. Li, J. Yuan, B. Tang, Z. Li, J. Luo, W. Wang, X. Hu, Q. Gong, and J. Wang, Nat. Phys. 20, 101 (2024).

[45] T. Ma and G. Shvets, New J. Phys. 18, 025012 (2016).

[46] F. Gao, H. Xue, Z. Yang, K. Lai, Y. Yu, X. Lin, Y. Chong, G. Shvets, and B. Zhang, Nat. Phys. 14, 140 (2017).

[47] Z. Zhang, Y. Tian, Y. Wang, S. Gao, Y. Cheng, X. Liu, and J. Christensen, Adv. Mater. 30, e1803229 (2018).

[48] Y. H. Ko and R. Magnusson, Adv. Opt. Mater. 10, 2200262 (2022).

[49] K. Y. Lee, S. Yoon, S. H. Song, and J. W. Yoon, Sci. Adv. 8, eadd8349 (2022).

[50] H. Choi, S. Kim, M. Scherrer, K. Moselund, and C.-W. Lee, Nanomaterials 13, 3152 (2023).

[51] K. Y. Lee, K. W. Yoo, Y. Choi, G. Kim, S. Cheon, J. W. Yoon, and S. H. Song, Nanophotonics 10, 1853 (2021).

[52] X. Zhang, H.-X. Wang, Z.-K. Lin, Y. Tian, B. Xie, M.-H. Lu, Y.-F. Chen, and J.-H. Jiang, Nat. Phys. 15, 582 (2019).

[53] A. A. Gorlach, D. V. Zhirihin, A. P. Slobozhanyuk, A. B. Khanikaev, and M. A. Gorlach, Phys. Rev. B 99, 205122 (2019).

[54] R. Jackiw and C. Rebbi, Phys. Rev. D 13, 3398 (1976).

[55] W. P. Su, J. R. Schrieffer, and A. J. Heeger, Phys. Rev. B 22, 2099 (1980).

[56] C. L. Dym and M. A. Lang, J. Acoust. Soc. Am. 56, 1523 (1974).

[57] See Supplemental Material at http://link.aps.org/supplemental/xx for (1) Band structure of acoustic guided mode resonance (2) Robustness analysis of acoustic Jackiw-Rebbi states (3) Two-dimensional generalization of acoustic Jackiw-Rebbi states (4) Details of numerical simulations.